\documentclass[preprint,final,5p,times,twocolumn]{elsarticle}
\usepackage{amssymb}
\usepackage{graphicx}
\usepackage{multirow}
\usepackage{color}
\usepackage{amsthm}
\usepackage{amsmath}
\usepackage{widetext}
\usepackage{amssymb}
\usepackage{float}
\usepackage[utf8x]{inputenc}
\usepackage[numbers]{natbib}
\usepackage[colorlinks=true,citecolor=blue]{hyperref}
\journal{Physics Letters B}
\usepackage[normalem]{ulem}

\begin{document}
\begin{frontmatter}
\title{Dirac vs. Majorana Dark Matter Imprints on Neutron Star Observables}
\author{M. Bhuyan$^{a}$}
\ead{mrutunjaya.b@iopb.res.in}
\author{Jeet Amrit Pattnaik$^{a}$}
\ead{jeetamritboudh@gmail.com}
\author{S. K. Patra$^{b}$}
\ead{sureshpatra@soa.ac.in}
\author{Sudhanwa Patra$^{a,c}$}
\ead{sudhanwa@iitbhilai.ac.in}
\address[1]{Institute of Physics, Sachivalya Marg, Bhubaneswar-751005, India}
\address[2]{Department of Physics, Siksha $'O'$ Anusandhan, Deemed to be University, Bhubaneswar-751030, India}
\address[3]{Department of Physics, Indian Institute of Technology Bhilai, Durg-491001, India}

\begin{abstract}
The fundamental character of a fermionic dark matter, whether it is a Dirac or Majorana particle-remains a key unresolved issue
whose answer would profoundly affect dark-sector phenomenology and detection strategies thereby motivates complementary probes across particle and astrophysical experiments. Compact stars, particularly neutron stars, offer unique astrophysical laboratories for probing such fundamental properties under extreme densities. The presence of a fermionic DM admixed with nuclear matter can modify the equation of state, thereby affecting observable quantities such as the mass–radius ($M-R$) relation and tidal deformability. In this work, we investigate how the intrinsic particle nature of fermionic DM influences neutron star structure. Within a relativistic mean-field framework extended by a scalar (or Higgs-like) portal coupling between DM and nucleons, we construct self-consistent equation of states for both Dirac and Majorana cases and solve the Tolman–Oppenheimer–Volkoff equations to obtain stellar configurations. Owing to the difference in internal degrees of freedom, Dirac DM ($g_{\chi}=4$) generally softens the equation of state more strongly than Majorana DM ($g_{\chi}=2$), leading to smaller radii and lower maximum masses. We identify the parameter space consistent with current NICER and gravitational-wave constraints, highlighting the potential of compact-star observations to discriminate between Dirac and Majorana dark matter.
\end{abstract}
\end{frontmatter}
\section{Introduction} \label{sec1}
The microscopic nature of dark matter (DM) remains one of the most profound open questions in modern particle physics and cosmology. Despite compelling astrophysical and cosmological evidence for its existence supported by a wide range of gravitational observations ranging from galactic rotation curves, gravitational lensing, to large-scale structure formation and cosmic microwave background (CMB) anisotropies, the particle identity of DM continues to elude detection \cite{Bertone2005, Planck2020, Freese2017}. A massive, electrically neutral fermion constitutes one of the simplest theoretical possibilities, yet it is unclear whether the fermionic dark matter is Dirac or Majorana in nature. In the case of Dirac dark matter, it possesses distinct particle and antiparticle degrees of freedom, while for the Majorana candidate, which is self-conjugate, implying particle identical to its antiparticle. Both scenarios are compatible with the existence of a massive, neutral fermion, but they differ in their symmetry properties, interactions, and thermodynamic behaviour \cite{Cirelli2011, Arcadi2020}. While direct detection, indirect searches, and collider experiments probe the couplings and masses of DM candidates, distinguishing between Dirac and Majorana fermions in terrestrial experiments is challenging, especially for weakly interacting scenarios where cross sections are immensely suppressed \cite{Schumann2019, Bottino2003, Essig2012}. This has motivated complementary approaches that exploit astrophysical environments—where extreme densities and gravitational fields can amplify otherwise negligible DM effects.

In this context, compact stars, particularly neutron stars (NSs) and hybrid stars with quark cores, may provide a unique astrophysical laboratory for probing the nature of fermionic DM under extreme conditions. Stellar structure is governed by the equation of state (EoS) of dense matter, which relates pressure and energy density. The inclusion of an additional fermionic component, such as DM, modifies the EoS and thereby alters global observables such as the stellar mass–radius ($M$–$R$) relation, maximum mass, and tidal deformability inferred from gravitational wave (GW) observations \cite{Abbott2017, GW170817, GW190814}. Recent high-precision measurements of NS radii by the NICER mission and the discovery of $\sim 2 M_\odot$ pulsars have imposed stringent constraints on any EoS modification, providing powerful tests for exotic components such as DM \cite{Miller2019, Miller2021, Riley2019, Antoniadis2013, Cromartie2020, Das2020, Dey2024, patt25}. Previous studies have investigated the generic impact of DM on compact-star properties, typically focusing on its gravitational or pressure effects without distinguishing its intrinsic particle nature \cite{Dey2024, patt25, Leung2011, Ciarcelluti2011, Panotopoulos2017}. These analyses generally show that DM softens the EoS, reducing the maximum mass and altering the tidal deformability. However, the Dirac versus Majorana distinction, which directly affects the internal degrees of freedom and hence the energy density and pressure of the DM sector, has received comparatively little attention in the context of neutron star physics.

In this work, we go beyond the generic analysis and demonstrate that the Dirac or Majorana character of fermionic DM leaves distinct and potentially testable imprints on neutron star observables. Within the framework of the relativistic mean-field (RMF) approach for dense hadronic matter (extended to hybrid configurations when relevant), we incorporate a fermionic DM component that interacts with baryons through a scalar (Higgs-like) portal coupling. The scalar channel is particularly motivated because it directly influences bulk thermodynamics by modifying effective masses and scalar densities, and it admits a natural connection to Higgs-portal ultraviolet (UV) completions \cite{Baek2012, Arcadi2019}. We compute the DM contributions to the energy density and pressure for both Dirac and Majorana cases, embed them self-consistently into the total EoS, and solve the Tolman–Oppenheimer–Volkoff (TOV) equations to obtain the resulting mass–radius relations. By comparing these predictions with current observational bounds from pulsar mass measurements, NICER radii, and GW data, we identify viable regions of DM parameter space and assess the observational prospects for discriminating between Dirac and Majorana dark matter through compact-star astrophysics.  \\
The paper is organized as follows: Section~\ref{formalism} presents the theoretical framework for the equation of state of a hybrid star comprising nuclear, quarkonic, and dark matter components. The results and their discussion are provided in Section~\ref{result}, followed by concluding remarks in Section~\ref{summary}.
\section{Theoretical Frameworks} 
\label{formalism}
We consider an effective low-energy Lagrangian that simultaneously captures the dynamics of nucleonic matter within the Relativistic Mean-Field (RMF) framework \cite{Serot1986, Ring1996, Dutra2014} and incorporates the emerging quarkonic matter concept proposed in Refs \cite{Dey2024, patt25, McLerran2019, Duarte2020,Sen2021}. At higher baryon densities, quarks begin to populate the low-momentum states of the Fermi sea, while nucleons remain confined near the Fermi surface. This framework preserves colour confinement at the surface of Fermi spheres while allowing quark degrees of freedom to dominate bulk thermodynamics at ultra-nuclear densities. It thus provides a smooth interpolation between the hadronic and deconfined quark phases, avoiding the necessity of a sharp first-order phase transition. To further extend this framework to include the dark sector, we augment the Lagrangian with a fermionic dark matter (DM) field $\chi$, which interacts with baryons via a scalar mediator field. For simplicity, the mediator is taken to be either the Standard Model (SM) Higgs boson or a Higgs-like scalar emerging from extensions of the SM \cite{Arcadi2020, Cline2013}. The relevant Lagrangian is given by \cite{Dey2024, pattcurv25}, \begin{eqnarray}
{\cal L} = {\cal L_{NM}} + {\cal L_{QM}} + {\cal L_{\chi}}. 
\end{eqnarray}
Here ${\cal L_{NM}}$, ${\cal L_{QM}}$, and ${\cal L_{\chi}}$ stands for the corresponding Lagrangian of Nucleonic, Quarkonic, and dark matter, respectively. The details are given in the following sub-sections. Using the above Largangian, the effective masses, and subsequently the energy density and pressure of the system, are determined self-consistently through the coupled field equations. The resulting equation of state (EoS) reflects the interplay among nuclear, quarkonic, and dark components, and is sensitive to the strength of the scalar coupling and the mediator mass \cite{Dey2024,Ellis2018}.
\subsection{Nuclear Model} \label{a}
Relativistic Mean Field (RMF) theory offers a self-consistent and unified framework to describe dense baryonic matter from finite nuclei to neutron star interiors—through meson-mediated nucleon interactions. It successfully reproduces empirical nuclear saturation properties and fiinite-nuclei observables \cite{Lalazissis1997,Singh2013,Estal2004,Lourenco2021}. In this work, we employ the effective RMF (E-RMF) model \cite{Singh2013,Todd-Rutel2005}, which extends the conventional RMF formalism by including nonlinear self- and cross-couplings among the $\sigma$, $\omega$, $\rho$, and $\delta$ mesons. Leptons are incorporated as relativistic free Fermi gases. The total energy density ($\mathcal{E}_{\rm NML}$) and pressure ($P_{\rm NML}$) of $\beta$-equilibrated matter are obtained from the energy–momentum tensor following Refs. \cite{Singh2013,G3,IOPB,patt22}, 
\begin{eqnarray}
\label{eq:eden}
{\cal E}_{\rm NML} & = & \sum_{i=p,n} \frac{g_s}{(2\pi)^{3}}\int_{0}^{k_{f_{i}}} d^{3}k\, \sqrt{k^{2} + M_{\rm nucl.}^{*2}}\nonumber\\
& + & n_{b} g_\omega\,\omega+m_{\sigma}^2{\sigma}^2\Bigg(\frac{1}{2}+\frac{\kappa_{3}}{3!}\frac{g_\sigma\sigma}{M_{\rm nucl.}}+\frac{\kappa_4}{4!}\frac{g_\sigma^2\sigma^2}{M_{\rm nucl.}^2}\Bigg)
\nonumber\\
&-&\frac{1}{4!}\zeta_{0}\,{g_{\omega}^2}\,\omega^4
 -\frac{1}{2}m_{\omega}^2\,\omega^2\Bigg(1+\eta_{1}\frac{g_\sigma\sigma}{M_{\rm nucl.}}+\frac{\eta_{2}}{2}\frac{g_\sigma^2\sigma^2}{M_{\rm nucl.}^2}\Bigg)
 \nonumber\\
& +& \frac{1}{2} (n_{n} - n_{p}) \,g_\rho\,\rho
 -\frac{1}{2}\Bigg(1+\frac{\eta_{\rho}g_\sigma\sigma}{M_{\rm nucl.}}\Bigg)m_{\rho}^2
 \nonumber\\
 & -& \Lambda_{\omega}\, g_\rho^2\, g_\omega^2\, \rho^2\, \omega^2
+\frac{1}{2}m_{\delta}^2\, \delta^{2}
\nonumber\\
&+&\sum_{j=e,\mu}  \frac{g_s}{(2\pi)^{3}}\int_{0}^{k_{f_{j}}} \sqrt{k^2 + m^2_{j}} \, d^{3}k,
\end{eqnarray}

\begin{eqnarray}
\label{eq:press}
P_{\rm NML} & = & \sum_{i=p,n} \frac{g_s}{3 (2\pi)^{3}}\int_{0}^{k_{f_{i}}} d^{3}k\, \frac{k^2}{\sqrt{k^{2} + M_{\rm nucl.}^{*2}}} \nonumber\\
& - & m_{\sigma}^2{\sigma}^2\Bigg(\frac{1}{2} + \frac{\kappa_{3}}{3!}\frac{g_\sigma\sigma}{M_{\rm nucl.}} + \frac{\kappa_4}{4!}\frac{g_\sigma^2\sigma^2}{M_{\rm nucl.}^2}\Bigg)+ \frac{1}{4!}\zeta_{0}\,{g_{\omega}^2}\,\omega^4 
\nonumber\\
& +& \frac{1}{2}m_{\omega}^2\omega^2\Bigg(1+\eta_{1}\frac{g_\sigma\sigma}{M_{\rm nucl.}}+\frac{\eta_{2}}{2}\frac{g_\sigma^2\sigma^2}{M_{\rm nucl.}^2}\Bigg)
\nonumber\\
&+& \frac{1}{2}\Bigg(1+\frac{\eta_{\rho}g_\sigma\sigma}{M_{\rm nucl.}}\Bigg)m_{\rho}^2\,\rho^{2}-\frac{1}{2}m_{\delta}^2\, \delta^{2}+\Lambda_{\omega} g_\rho^2 g_\omega^2 \rho^2 \omega^2
\nonumber\\
& + & \sum_{j=e,\mu}  \frac{g_s}{3(2\pi)^{3}}\int_{0}^{k_{f_{j}}} \frac{k^2}{\sqrt{k^2 + m^2_{j}}} \, d^{3}k.
\end{eqnarray}
The different mesonic masses and couplings used here follow Refs.~\cite{Singh2013,G3,IOPB,patt22}.
\subsection{Quarkyonic Model} \label{b}
At densities well above saturation, baryons begin to reveal quark substructure. The quarkyonic matter concept \cite{McLerran2019,Duarte2020,Sen2021} assumes quarks populate low-momentum states while nucleons remain confined near the Fermi surface. A rapid stiffening of the EoS near the transition density is a key signature. Following ReF. \cite{Duarte2020}, the system simultaneously satisfies baryon-number conservation, charge neutrality, and $\beta$-equilibrium, leading to \cite{Dey2024,patt25,pattcurv25,Dey_fmode}:
\begin{eqnarray}
n &=& n_n + n_p + \frac{n_u+n_d}{3}
\nonumber\\
&=& \frac{g_s}{6\pi^2}\bigg[(k_{f_{n}}^3-k_{0_{n}}^3)+(k_{f_{p}}^3-k_{0_{p}}^3)+\frac{(k_{f_{u}}^3+k_{f_{d}}^3)}{3}\bigg],
\end{eqnarray}
Charge neutrality is enforced through:
\begin{eqnarray}
n_p + \frac{2n_{u}}{3} -  \frac{n_{d}}{3}= n_{e^{-}} + n_{\mu}.
\end{eqnarray}
The minimum momenta for nucleons relate to the transition momentum $k_{\rm Ft}$ by \cite{Duarte2020}:
\begin{eqnarray}
 k_{0(n,p)} &=& (k_{f_{(n,p)}}-k_{t_{(n,p)}})\bigg[1+ \frac{\Lambda_{\rm cs}^2}{k_{f_{(n,p)}}k_{t_{(n,p)}}}\bigg].
\end{eqnarray}
Chemical equilibrium among neutrons, protons, and quarks requires:
\begin{eqnarray}\label{ebnq}
\mu_n &=& \mu_u + 2\mu_d, \\
\mu_p &=& 2\mu_u + \mu_d.
\end{eqnarray}
The weak interaction imposes beta-equilibrium:
\begin{eqnarray} \label{qnbe}
\mu_{n} &=& \mu_{p} + \mu_{e^{-}}, \nonumber \\
\mu_{\mu} &=& \mu_{e^{-}}.
\end{eqnarray}
The light-quark masses at onset follow:
\begin{eqnarray}
m_u &=& \frac{2}{3} \mu_{t_p} - \frac{1}{3} \mu_{t_n}, \nonumber \\
m_d &=& \frac{2}{3} \mu_{t_n} - \frac{1}{3} \mu_{t_p}.
\end{eqnarray}
Quarks are treated as a non-interacting Fermi gas, giving:
\begin{eqnarray}
{\cal E}_{\rm QM}&=& \sum_{j=u,d}\frac{g_s N_c}{(2\pi)^3}\int_0^{k_{f_{j}}}k^2\sqrt{k^2 + m_{j}^2 }\, d^3k,
\end{eqnarray}
\begin{eqnarray}
P_{\rm QM} &=& \mu_{u} n_{u} + \mu_{d} n_{d} - \epsilon_{QM}.
\end{eqnarray}
\subsection{Dark Matter Model} 
\label{c}
We now intend to write down the interaction Lagrangian relevant for both Dirac and Majorana fermionic dark matter. The fermionic dark matter (DM) interacting through a Higgs-like scalar portal provides a well-motivated framework for studying DM effects in neutron star interiors. Here, the fermionic dark fermion $\chi$ couples to a scalar mediator $h$ that also interacts with nucleons, thereby modifying the in-medium effective masses of both sectors. The corresponding DM Lagrangian takes the form, 
\begin{equation}
\mathcal{L}_{\chi} =
\begin{cases}
\bar{\chi}\left(i\gamma^{\mu}\partial_{\mu} - M_{\chi} + y_{\chi} h\right)\chi, & \text{Dirac DM}, \\[6pt]
\frac{1}{2}\bar{\chi}\left(i\gamma^{\mu}\partial_{\mu} - M_{\chi} + y_{\chi}h\right)\chi, & \text{Majorana DM}.
\end{cases}
\end{equation}
Here, the additional factor of $\frac{1}{2}$ signifies the self-conjugate nature of the Majorana DM. The effect of dark matter using a scalar portal in a neutron star has been explored in recent works without distinguishing their (i.e., Dirac or Majorana) nature \cite{Arcadi2020, Das2020, Dey2024, patt25, Cline2013, pattcurv25, Dey_fmode}. Using the mean-field approximation, the DM energy density and pressure become:
\begin{eqnarray}
{\cal{E}}_{\chi}& = & \frac{g_{\chi}}{(2\pi)^{3}}\int_0^{k_f^{\chi}} d^{3}k \sqrt{k^2 + (M_\chi^\star)^2} + \frac{1}{2}M_h^2 h_0^2 ,
\end{eqnarray}
\begin{eqnarray}
P_{\chi}& = &\frac{g_{\chi}}{3(2\pi)^{3}}\int_0^{k_f^{\rm DM}} \frac{d^{3}k k^2} {\sqrt{k^2 + (M_\chi^\star)^2}} - \frac{1}{2}M_h^2 h_0^2 .
\end{eqnarray}
The degeneracy factor $g_{\chi} =4$ for a Dirac fermion DM and $g_{\chi}=2$ for a Majorana fermion DM. Accordingly, the number density is
\begin{equation}
n_{\chi} = \frac{g_{\chi}}{6\pi^2}\,(\rm k_{\rm f}^{\rm DM})^3.
\label{eq:nchi} 
\end{equation}
Note that for the same $\rm k_{\rm f}^{\chi}$, the Dirac case carries twice the degrees of freedom of the Majorana case, and thus contributes roughly twice the pressure and energy density in the ultrarelativistic or non-relativistic limits. 

The effective masses of nucleons and DM modified by Higgs mediation are:
\begin{eqnarray}
M_i^\star &=& M_{\rm nucl.} + g_\sigma \sigma_0 \mp g_\delta \delta_0 - \frac{f M_{\rm nucl.}}{v} h_0, 
\nonumber\\
M_\chi^\star &=& M_\chi - y h_0.
\end{eqnarray}
Noting the DM contribution to enegry and pressure density scale linearly with $g_\chi$. As a result, Dirac fermionic dark matter soften the EoS of neutron star in comparison to Majorana contributions for identical dark sector mass and couplings. These differences in effective mass, energy and pressure density may result in distinct modifications to the mass-radius trajectory and tidal behaviour. Thus neutron star observables including recent observations \cite{Abbott2017,GW170817,GW190814} can potentially discriminate between the Dirac and Majorana nature of DM. 

\section{Results and Discussion} \label{result}
Now including the DM contributions, the total thermodynamic quantities for dark matter–admixed quarkyonic matter given as:
\begin{eqnarray}
\cal{E} &=& {\cal{E}}_{\rm BM} + {\cal{E}}_{\rm QM} + {\cal{E}}_{\chi},
\label{etot}
\end{eqnarray}
\begin{eqnarray}
P &=& P_{\rm BM} + P_{\rm QM} + P_{\chi}.
\label{ptot}
\end{eqnarray}
Using Eq. (\ref{etot}), and (\ref{ptot}), we perform a systematic exploration of dark matter admixed quarkonic stars (DAQSs) with particular attention to the intrinsic nature of the dark matter (DM), distinguishing between Dirac and Majorana types. The motivation is that the underlying spin and internal degrees of freedom of DM can substantially influence the dense matter equation of state (EoS), and consequently, the macroscopic observables of compact stars. In our setup, the quarkonic star configuration is defined by a transition density $\rm n_{\rm t}$, beyond which matter deconfines into a quark Fermi sea, characterised by the QCD confinement scale $\Lambda_{\rm cs}$. The resulting hybrid star, therefore, contains a nucleonic outer shell smoothly matched to a deconfined quark core in charge neutrality and $\beta$-equilibrium. For the hadronic sector, we employ two well-established parameterizations of the effective field theory–motivated relativistic mean-field (E-RMF) approach, namely G3 \cite{G3} and IOPB-I \cite{IOPB} parameter sets. These parametrizations are chosen for their consistency with recent nuclear matter constraints and astrophysical data. The total energy density and pressure of the system are modified by the inclusion of a DM Fermi sea, where the effect depends critically on whether the DM is Dirac (both spin and particle/antiparticle degrees of freedom) or Majorana DM (self-conjugate, contributes only spin degrees of freedom) \cite{Ellis2018,xiang2014}.
\begin{figure}
\includegraphics[width=1.0\columnwidth]{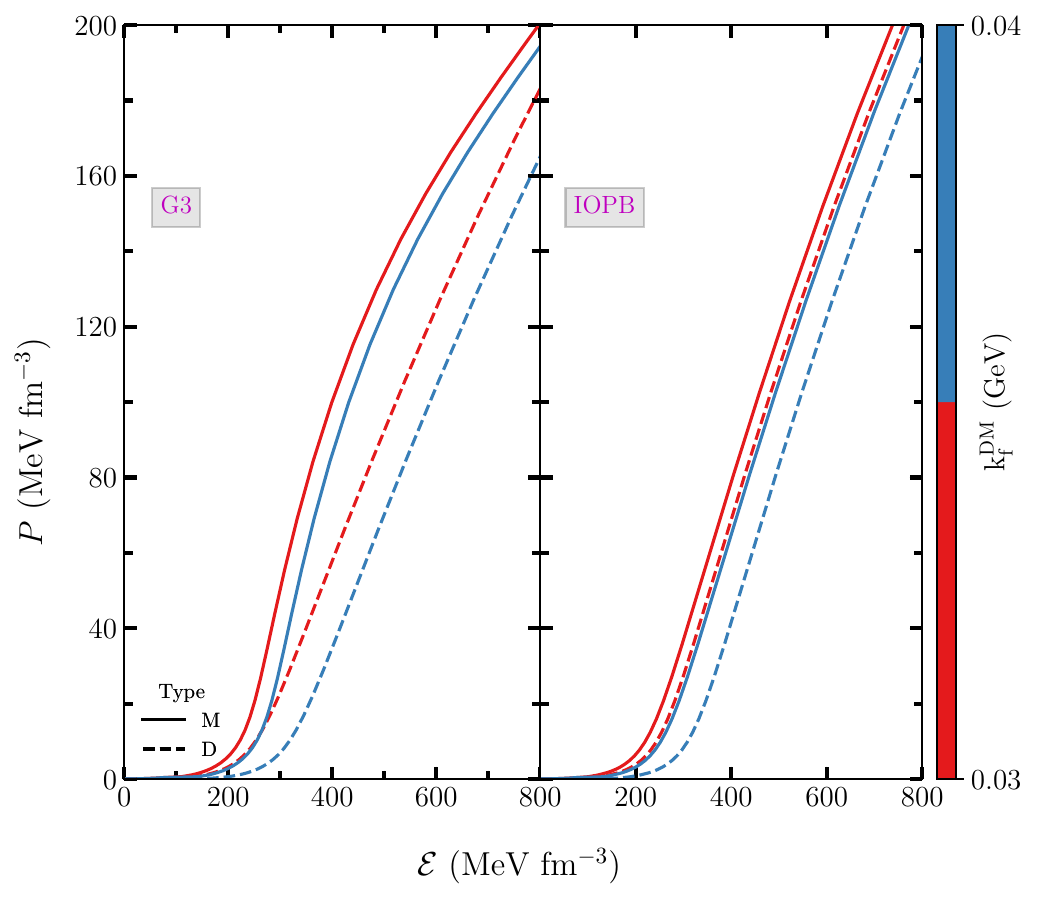}
\caption{Equation of state for dark matter–admixed quarkonic star having transition density $\rm n_{\rm t} = 0.3\,\,\mbox{fm}^{-3}$, QCD confinement scale $\Lambda_{\rm cs} = 800\, $ MeV and DM Fermi momentum $\rm k_{\rm f}^{\rm DM}=$ 0.03 GeV (red) and 0.04 GeV (blue) with the G3 and IOPB-I forces. The solid and dashed lines represent the Majorana and Dirac nature, respectively.}
\label{fig1}
\end{figure}

The resulting EoSs are plotted in Fig. \ref{fig1}, where the left and right panels correspond to the G3 and IOPB-I parametrizations, respectively. For each case, we consider two representative DM Fermi momenta, $\rm k_{\rm f}^{\rm DM}$ = 0.03 GeV and 0.04 GeV, at a fixed confinement scale of $\Lambda_{\rm cs}$ = 800 MeV. The EoSs with Dirac ($g_\chi = 4$) and Majorana ($g_\chi = 2$) DM are displayed as dotted and solid curves, respectively, with colour coding denoting the DM Fermi momenta. From these results, we note a clear stiffening of the EoS in the presence of Majorana DM relative to the Dirac case. This stiffening arises because Majorana DM carries $\frac{1}{2}$ of the internal degrees of freedom as compared to the Dirac case, thereby producing a softening effect on the pressure at high densities. Consequently, the maximum pressure achievable at a given energy density is higher in the Majorana case. 

\begin{figure}
\includegraphics[width=1.0\columnwidth]{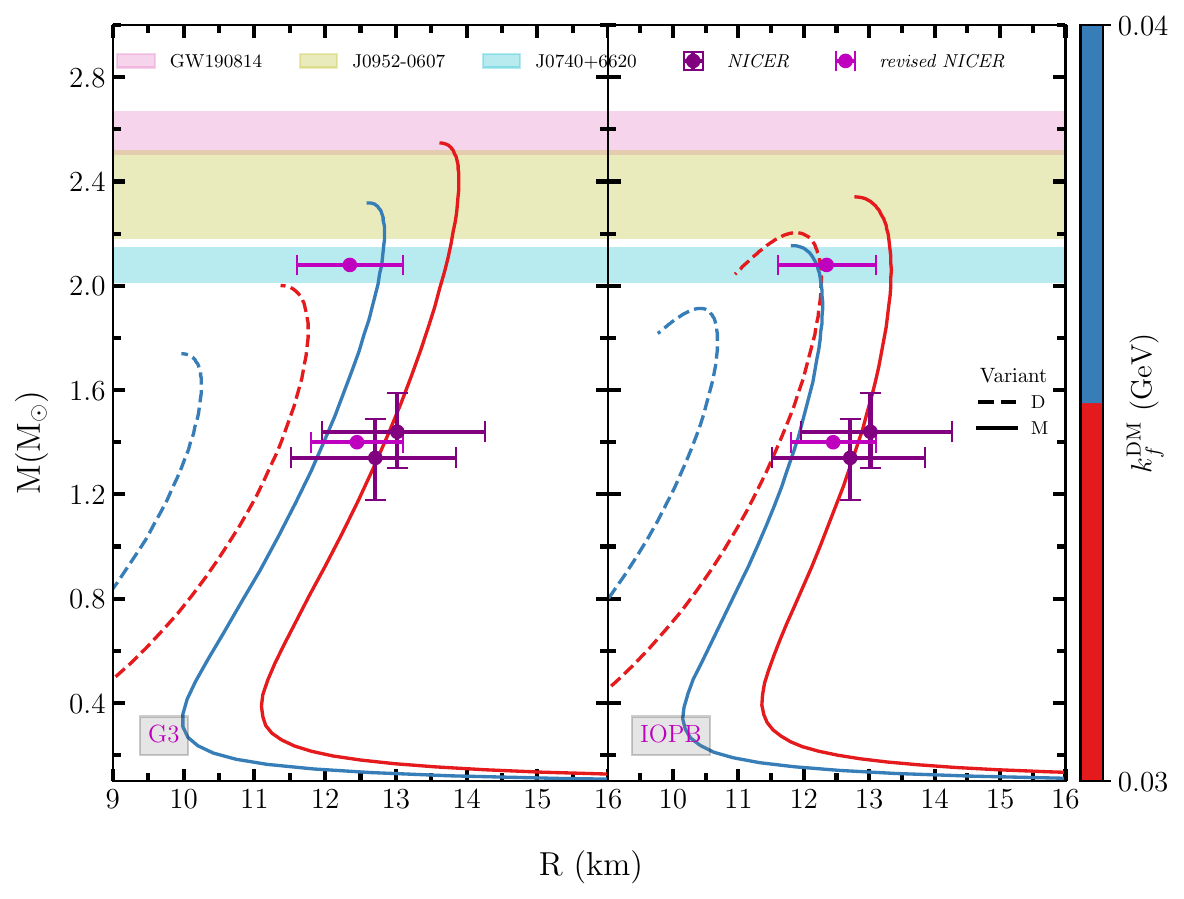}
\caption{Mass–radius relation of dark matter–admixed quarkonic stars obtained using the G3 and IOPB-I parameter sets with $\rm n_{\rm t} = 0.3\,\,\mbox{fm}^{-3}$. The solid (dashed) curve corresponds to Majorana (Dirac) dark matter contribution to $M-R$ relation. The horizontal shaded regions represent observational constraints from recent pulsar \cite{Miller2019,Miller2021} and gravitational wave data \cite{GW170817,Riley2019}.}
\label{fig2}
\end{figure}

The astrophysical implications for the behavior of EoSs are further illustrated by solving the Tolman–Oppenheimer–Volkoff (TOV) equations \cite{Tolman1939,Oppenheimer1939} using the calculated ${\cal E}$ in Eqn. (\ref{etot}) and $P$ in Eqn. (\ref{ptot}), 
\begin{align}
\frac{dP(r)}{dr} &= -\frac{G}{r^2}\,\frac{\big[{\cal E} (r)+P(r)\big]\big[M(r)+4\pi r^3 P(r)\big]}{1-2GM(r)/r}, \\
\frac{dM(r)}{dr} &= 4\pi r^2 {\cal E} (r),
\end{align}
with central boundary conditions $M(0)=0$ and $P(0)=P_c$. For chosen central values (equivalently, central baryon and DM densities), the integration proceeds outward until $P(R)=0$, determining the stellar radius $R$ and gravitational mass $M(R)$. As shown in Fig. \ref{fig2}, the corresponding mass–radius (M–R) relations are obtained for DM-admixed quarkonic stars systematically. From the figure, it is observed that the Dirac DM admix stars yield smaller radii and lower maximum masses. At the same time, the Majorana-admixed configurations remain consistent with observational constraints from NICER on PSR J0030+0451 and PSR J0740+6620 \cite{Riley2019}, as well as tidal deformability bounds inferred from GW170817 \cite{GW170817}. Our systematic scan demonstrates that, for $\rm k_{\rm f}^{\rm DM}$= 0.03 GeV, the Majorana DM admixed quarkonic EoSs simultaneously satisfy the NICER mass–radius constraints and the lower bound on the maximum neutron star mass ($\sim$ 2.4 $\rm M_{\odot}$. In contrast, the Dirac DM case underpredicts the radius and/or fails to meet the mass threshold. At  $\rm k_{\rm f}^{\rm DM}$=0.04 GeV, both Dirac and Majorana DM induce significant softening, pushing the stellar configurations outside the NICER limits, thereby disfavoring larger DM admixtures. Taken together, these results suggest that if DM is indeed present in quarkonic stars, its Majorana nature is strongly favoured. This conclusion is robust across both hadronic parameterisations employed in this study and remains stable against moderate variations in the transition density and confinement scale.
\begin{figure}
\includegraphics[width=1.0\columnwidth]{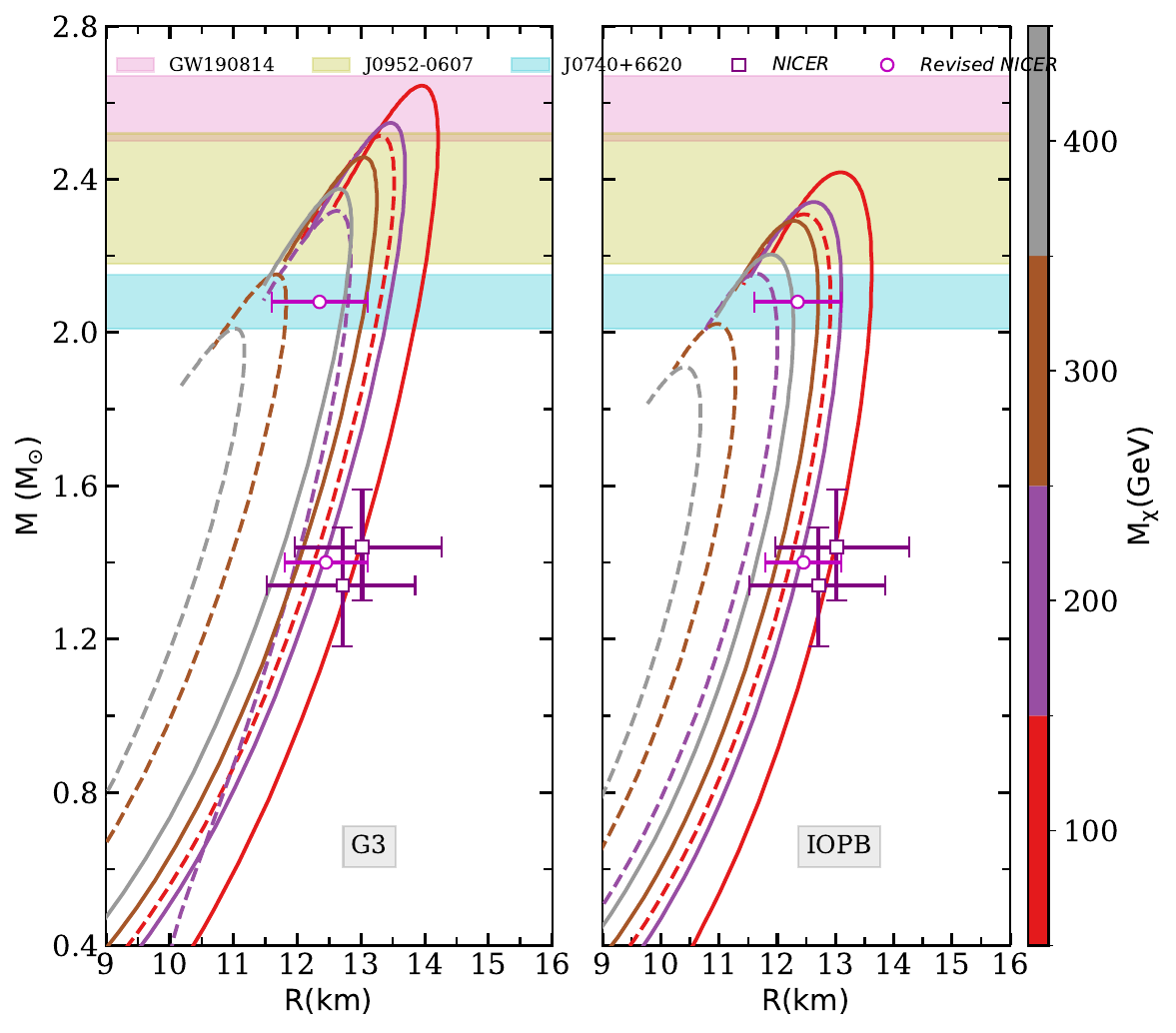}
\caption{M-R relations for the Majorona type EoSs having $\rm n_{\rm t} = 0.3\, {\rm fm}^{-3}$, QCD confinement scale $\Lambda_{\rm cs} = 800$ MeV and DM Fermi momentum $\rm k_{\rm f}^{\rm DM}$  = 0.03 GeV (solid) and 0.04 GeV (dashed) at different DM masses M$_{\chi}$ = 100, 200, 300, 400 GeV, for the G3 and IOPB-I forces, along with the observational data \cite{GW170817,Miller2019,Miller2021,Riley2019}.}
\label{fig3}
\end{figure}
\begin{figure}
\includegraphics[width=1.0\columnwidth]{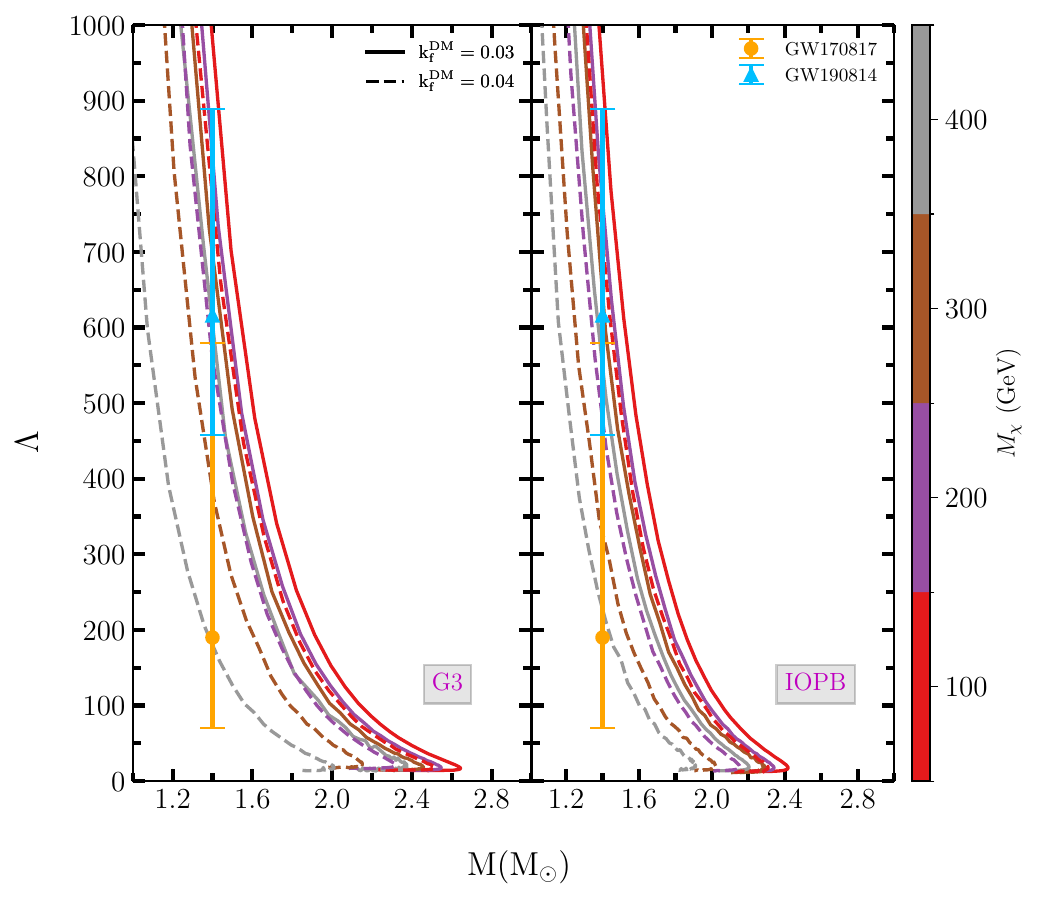}
\caption{$\Lambda$ - M relations for the Majorona type EoSs having $\rm n_{\rm t} = 0.3\, {\rm fm}^{-3}$, QCD confinement scale $\Lambda_{\rm cs} = $ 800 MeV and DM Fermi momentum $\rm k_{\rm f}^{\rm DM}=$ 0.03 GeV (solid) and 0.04 GeV (dashed) at different DM masses M$_{\chi}$ = 100, 200, 300, 400 GeV, for the G3 and IOPB-I forces, along with the observational data \cite{GW170817,Riley2019}.}
\label{fig4}
\end{figure}

In the above analysis, we found that the neutron star observables constraints in Refs. \cite{Riley2019,GW170817} are satisfied with the given set of benchmark parameters for Majorana type fermionic dark matter admixed quarkonic star. Thus, it is appropriate to investigate further the effect of variation of dark matter masses for EoS corresponding to the Majorana DM. The calculated mass–radius (M–R) relations for Majorana-type dark matter admixed quarkonic stars (DAQSs) while varying dark matter masses in the range $100-600$~GeV using the G3 and IOPB-I effective forces are shown in Fig. \ref{fig3}. The numerical results are obtained for same benchmark points including the transition density $\rm n_{\rm t} = 0.3 \rm fm^{-3}$, QCD confinement scale $\Lambda_{\rm cs} =$ 800 MeV while varying the Majorana DM masses. The solid and dashed curves as shown in Fig. \ref{fig3} are corresponding to the DM Fermi momentum values $\rm k_{\rm f}^{\rm DM} = $ 0.03 GeV and 0.04 GeV, respectively. Across both nuclear forces, the inclusion of DM systematically softens the equation of state (EoS), leading to a suppression of the maximum mass and radius as comparared to the scenario without dark matter. Quantitatively, for the G3 force, the maximum mass decreases from $\rm M_{max} \simeq 2.05 \rm M_{\odot}$ (no DM) to $1.85 \rm M_{\odot}$ at $\rm M_{\chi} = $400 GeV and $\rm k_{\rm f}^{\rm DM} = $ 0.04 GeV, while the corresponding radius at $1.4 \rm M_{\odot}$ decreases from $\sim 12.6$ km to $\sim 11.4$ km. In other hand, IOPB-I force shows a milder reduction, with $M_{\text{max}}$ decreasing from $\sim 2.15 \rm M_{\odot}$ to $1.95 \rm M_{\odot}$ over the same parameter range, and the radius at $1.4 \rm M_{\odot}$ suppressed from $\sim 13.0$ km to $\sim 11.8$ km. These shifts of $\Delta \rm M \approx 0.2 \rm M_{\odot}$ and $\Delta R \approx 1$–$1.2$ km demonstrate that even modest DM fractions can produce significant structural modifications. Importantly, at $\rm k_{\rm f}^{\rm DM} = $ 0.04 GeV, the predicted radii for $1.4 \rm M_{\odot}$ stars fall within the range 11.0–11.8 km, which is compatible with the NICER radius constraints for PSR J0030+0451 ($R_{1.4} = 12.71^{+1.14}_{-1.19}$ km) \cite{Miller2019} and PSR J0740+6620 ($R{2.08} = 12.39^{+1.30}_{-0.98}$ km) \cite{Miller2021}.

We now intend to examine how a neutron star or hybrid star get deformed when Majorana dark matter taken into consideration. The corresponding dimensionless tidal deformability ($\Lambda$) as a function of stellar mass is shown in Fig. \ref{fig4} for the same parameter sets.  In Fig. \ref{fig4}, the tidal deformability ($\Lambda$) with $1.4 \rm M_{\odot}$ stars has been estimated by varying the dark matter mass in the range $100-400$~GeV using both sets G3 (left panel) and IOPB-I (right-panel), respectively. The results are then compared with the present observational constraints, namely, GW170817 (orange curve) and GW190814 (green curve). It is observed from the Fig. \ref{fig4} that inclusion of Majorana DM significantly reduces $\Lambda$ across all masses, consistent with the usual compact configurations. For example, in case of $1.4 \rm M_{\odot}$ stars, the tidal deformability $\Lambda_{1.4}$ decreases from $\sim 500$ (IOPB-I parametrization without DM) to $\sim 350$ at $\rm M_{\chi} = $400 GeV and $\rm k_{\rm f}^{\rm DM} = $ 0.04 GeV, representing a 30\% reduction. While for the G3 case, the value of $\Lambda_{1.4}$ decreases from $\sim 450$ to $\sim 310$ over the same DM parameter space. Notably, these reduced $\Lambda_{1.4}$ values fall well within the GW170817 constraint $\Lambda_{1.4} = 190^{+390}_{-120}$, while the DM-free models tend to overshoot the observational upper bound. At higher stellar masses ($M \gtrsim 2.0 M{\odot}$), the impact of DM is even more pronounced, with $\Lambda$ values reduced by up to 40\% compared to the DM-free case, reflecting the strong central concentration of DM in massive stars.

Taken together, these results highlight two key effects of Majorana-type DM on compact stars: (i) heavier DM particle masses and larger DM Fermi momenta lead to stronger EoS softening, producing systematically smaller radii ($\Delta R \sim 1$ km at $1.4 \rm M_{\odot}$) and tidal deformabilities (20–40\% suppression of $\Lambda_{1.4}$), and (ii) the degree of modification is sensitive to the nuclear force model, with IOPB-I producing stiffer configurations that remain consistent with both the 2.0 $\rm M_{\odot}$ pulsar mass constraint and GW170817 tidal deformability limits when DM is included. Thus, the combined M–R and $\Lambda$–M analysis favors the presence of Majorana-type DM with mass $\rm M_{\chi} \sim$ 400 GeV and fermi momentum $\rm k_{\rm f}^{\rm DM} \sim$ 0.03–0.04 GeV can reconcile theoretical EoSs with current astrophysical observations, GW170817 and GW190814 data, offering an indirect probe of DM properties through multimessenger astronomy. Finally, in comparison to Dirac DM, we note that Majorana particles' internal degrees of freedom generally produce stiffer EoSs for the same $\rm M_{\chi}$ and $\rm k_{\rm f}^{\rm DM}$, leading to slightly larger maximum masses and tidal deformabilities. Therefore, if any future observations find support for smaller radii and $\Lambda$ values, they may hint towards the Dirac nature of DM. In contrast, results consistent with stiffer configurations could be more compatible with Majorana-type DM.

\section{Conclusion}
\label{summary}
We have carried out a systematic investigation of the impact of fermionic dark matter on the structure of the quarkonic stars, emphasizing the distinction between Dirac and Majorana DM. Within the effective relativistic mean-field (E-RMF) framework using the G3 and IOPB-I parameter sets, we modeled DM as a cold Fermi sea coupled to baryons through a scalar (Higgs-like) portal. Our analysis shows that the inclusion of DM softens the EoS, leading to reduced stellar masses and radii. Quantitatively, the maximum mass decreases by about $\Delta M \sim 0.2 \rm M_{\odot}$, while the corresponding radius contracts by $\Delta R \sim 1–1.2$ km for typical DM admixtures. For instance, in the G3 model, the maximum mass drops from $\sim 2.05 \rm  M_{\odot}$ without DM to $\sim 1.85 \rm M_{\odot}$ at $\rm M_{\chi} = $400 GeV and $\rm k_{\rm f}^{\rm DM} = $ 0.04 GeV, with the radius at $1.4 \rm M_{\odot}$ shrinking from $\sim 12.6$ km to $\sim 11.4$ km. A similar but milder reduction is observed for IOPB-I. By solving the Tolman–Oppenheimer–Volkoff equations with these EoSs and comparing the results against current multimessenger constraints (NICER, $2 \rm M_{\odot}$ pulsars, and LIGO–Virgo–KAGRA tidal deformabilities), we delineate an allowed parameter space in $(\rm M_{\chi}, g_{\chi N}, n_{\chi}/n_{B})$ where the two DM hypotheses, Dirac versus Majorana, can be observationally distinguished.
A key outcome of this study is that Majorana DM yields a stiffer EoS than Dirac DM owing to its reduced internal degrees of freedom. Consequently, Majorana-admixed stars sustain larger radii and higher maximum masses for the same DM parameters. At $\rm k_{\rm f}^{\mathrm{DM}} = $ 0.03 GeV, Majorana-admixed stars remain consistent with both NICER and gravitational wave constraints. In contrast, Dirac-admixed stars often underpredict the radius or fail to satisfy the $2.0 \rm M_{\odot}$ mass limit. However, higher DM admixtures ($\rm k_{\rm f}^{\mathrm{DM}} \geq 0.04$ GeV) over-soften the EoS in both cases, disfavoring such scenarios. 

Our results indicate that Majorana DM with $\rm M_{\chi} \sim$400 GeV and $\rm k_{\rm f}^{\mathrm{DM}} \sim$0.03 GeV provides the best agreement with multimessenger data, whereas Dirac DM remains more tightly constrained. Further, it is found that the tidal deformability $\Lambda_{1.4}$ also decreased by 20–40 \%, bringing the predictions in agreement with the GW170817 and GW190814 constraints. In contrast, the EoSs without DM predictions tend to exceed the observational bounds. Overall, our findings suggested that precision measurements of mass, radius, and tidal deformability from NICER, LIGO–Virgo–KAGRA, and next-generation observatories could offer a novel astrophysical probe of the Dirac or Majorana nature of dark matter. This property has so far eluded terrestrial detection.
\section{Acknowledgement}
MB acknowledge the financial support under the Anusandhan National Research Foundation of Ramanujan Fellowship, File No. RJF/2022/000140. SP acknowledges the financial support under MTR/2023/000687 funded by SERB, Govt. of India.  


\end{document}